\documentclass[aps,pre,preprint,superscriptaddress,nofootinbib]{revtex4-2}
\usepackage[T1]{fontenc}
\usepackage{lmodern}
\usepackage{microtype}
\usepackage{amsmath,amssymb,mathtools}
\usepackage{booktabs}
\usepackage{graphicx}
\usepackage{xcolor}
\usepackage{hyperref}
\hypersetup{colorlinks=true,linkcolor=blue,citecolor=blue,urlcolor=blue}

\newcommand{\avg}[1]{\left\langle #1\right\rangle}
\newcommand{\eps}{\varepsilon}
\newcommand{\dd}{\mathrm{d}}
\newcommand{\vect}[1]{\mathbf{#1}}
\newcommand{\Ptot}{\vect P_{\rm total}}

\begin{document}

\title{Covariance-Driven Momentum Rectification at Liquid--Vapor Interfaces Near Wetting Transitions}
\author{Nelson Bol\'ivar}
\email{nelson.e.bolivar@ucv.ve}
\affiliation{Astrum Drive Technologies, Dallas Pkwy Unit 120 B, Frisco, Texas 75034, USA}
\affiliation{Departamento de F\'isica, Facultad de Ciencias, Universidad Central de Venezuela, Av.\ Los Ilustres, Caracas 1041-A, Venezuela}
\author{Gabriel Abell\'an}
\affiliation{Astrum Drive Technologies, Dallas Pkwy Unit 120 B, Frisco, Texas 75034, USA}
\affiliation{Departamento de F\'isica, Facultad de Ciencias, Universidad Central de Venezuela, Av.\ Los Ilustres, Caracas 1041-A, Venezuela}
\author{Ivaylo Vasilev}
\affiliation{Astrum Drive Technologies, Dallas Pkwy Unit 120 B, Frisco, Texas 75034, USA}
\date{\today}

\begin{abstract}
Zero-mean forcing can generate directed transport when a medium responds in a spatially structured way and the relevant symmetries are broken. Liquid--vapor interfaces are a useful setting for this problem because surface-tension gradients, wetting dynamics, vapor exchange, capillary and acoustic waves, electro-ionic screening, thermal noise, and boundary compliance can all carry momentum. We develop a conservative continuum model in which the central object is the covariance between a local response field and a zero-mean tangential drive, $\avg{Mf}-\avg{M}\avg{f}$, embedded in an explicit momentum ledger. In a minimal diffuse-interface realization, this covariance gives the phase-selective drift law $U\propto \epsilon_h\Theta\sin\varphi$ for a Marangoni-driven liquid--vapor interface above a structured wall, with exact nulls when the symmetry is restored. Wetting susceptibility then acts as a bounded gain factor: first-order spinodal conditions provide the useful amplification regime, critical wetting saturates because interfacial unbinding removes the short-range drive, and bulk criticality suppresses the channel as the interface disappears. Additional reservoirs - phase change, waves, thermal transport, electro-ionic coupling, compliance, and fluctuations - are treated as compensating channels to be isolated by sign reversals, scaling laws, and budget closure. The result is neither a new microscopic force nor an apparatus-level claim, but a symmetry-constrained accounting scheme for rectified momentum transfer in water-based interfacial systems. Reduced numerical calculations illustrate the phase-selection rules, bounded spinodal gain, momentum-budget closure, transverse chirality, and grid convergence.
\end{abstract}

\maketitle

\section{Introduction}

Directed motion from unbiased forcing is a standard result of nonequilibrium physics. Brownian motors and deterministic ratchets show how a zero-mean drive can produce a current when detailed balance, reflection symmetry, or time-reversal symmetry is broken \cite{Blanter1998,Reimann2002}. Hydrodynamic analogues appear in structured or anisotropic surfaces, tensorial slip, thermocapillary flows, thermally or chemically patterned films, Marangoni--Benard ratchets, superhydrophobic-surface thermocapillary slip, dielectrowetting, Leidenfrost droplets, soft wetting, and driven colloids \cite{Bazant2008,Zhou2012,Kataoka1999,Darhuber2003,Stroock2003,Baier2010,Yariv2018,YarivCrowdy2019,Crowdy2023,Linke2006,Ferraro2010,Edwards2018,Andreotti2020}. These literatures have different vocabularies, but they share a common mathematical feature: the average of a product is not generally the product of the averages. A drive with zero spatial mean can have a nonzero overlap with a structured response.

Liquid--vapor interfaces are a useful place to make this statement precise. They are not passive boundaries. In water and other volatile liquids, an interface can move, deform, exchange mass and heat, radiate capillary or acoustic waves, store surface energy, interact with ionic or dipolar structure, and exchange momentum with a wall or substrate. A measured net drift or force in such a system is therefore not interpreted by asking only whether the applied forcing has zero mean. One must also ask which reservoir carries the compensating momentum and whether the relevant symmetry operations produce the expected nulls and reversals.

This paper is best read as a conservative accounting model, not as a broad new transport claim. The starting point is deliberately conservative: if a subsystem acquires net momentum, the opposite momentum must be assigned to another part of the enlarged system. We therefore treat prior work as a constraint from the outset. Thermocapillary microflows, Marangoni ratchets, transverse flow over grooved superhydrophobic surfaces, and asymmetric thermocapillary pumping are established topics \cite{Kataoka1999,Darhuber2003,Stroock2003,Baier2010,YarivCrowdy2019,Crowdy2023}. The claim made here is narrower: covariance supplies the selection rule, wetting susceptibility controls the bounded gain, and momentum conservation identifies the reservoirs that must be characterized before any measurement is extrapolated beyond its tested regime.

This distinction is important for both theory and experiment. If the relevant reservoir is a wall, the effect is boundary-mediated. If it is vapor, the effect is phase-change-mediated. If it is a wave field, the signal should depend on acoustic or capillary resonances. If it is an electro-ionic channel, screening length and salinity should matter. If no such channel can be identified, the result is not automatically new physics; it is an incomplete model. The useful outcome, in either case, is a sharper problem: a channel with quantitative signatures, or a residual with named exclusions.

The paper proceeds as follows. Section 2 derives the momentum ledger from translation symmetry and rewrites it in continuum form. Section 3 introduces covariance-driven rectification at a liquid--vapor interface and the associated Marangoni drift law. Section 4 shows how wetting susceptibility creates bounded gain and explains why first-order spinodal, critical wetting, and bulk criticality behave differently. Section 5 extends the ledger to phase change, waves, thermal, electro-ionic, compliance, and stochastic channels, and states the falsification logic. Section 6 summarizes the claim discipline and the conclusions. The appendix describes the reduced numerical calculations and consistency checks used to test the manuscript-level predictions.

\section{Momentum accounting in a continuum medium}

The relevant constraint is not Newton's third law in its elementary pairwise form, but the conservation law that follows from spatial translation symmetry. For a closed system in a translation-invariant environment,
\begin{equation}
\frac{\dd \Ptot}{\dd t}=0.
\label{eq:global_conservation}
\end{equation}
If one part of the system gains momentum, another part must carry the opposite amount. The compensating term may be a material mass, but it may also be a field, wave, wall, interface, thermal flux, or another degree of freedom eliminated from an effective description.

This distinction matters because fluid mechanics is already a coarse-grained theory. A water molecule is not tracked individually. Instead one describes density $\rho(\vect x,t)$, velocity $\vect v(\vect x,t)$, pressure $p$, temperature $T$, chemical potential $\mu$, phase/order parameter $\phi$, and other fields. Coarse graining does not erase translation symmetry. It changes its form. Momentum conservation becomes a local balance between storage, flux through boundaries, and explicitly external forces:
\begin{equation}
\frac{\dd}{\dd t}\int_V g_i\,\dd V
= -\oint_{\partial V}\Pi_{ij}n_j\,\dd A
+\int_V f_i^{\rm ext}\,\dd V .
\label{eq:control_volume}
\end{equation}
Here $g_i$ is momentum density and $\Pi_{ij}$ is the momentum flux or stress tensor. In a simple fluid $g_i=\rho v_i$. The flux tensor contains advective momentum transport and stress; for an inviscid fluid one may write schematically
\begin{equation}
\Pi_{ij}=\rho v_i v_j+p\delta_{ij},
\end{equation}
while viscosity adds dissipative stress. Pressure and viscosity are therefore not add-ons to the accounting. They are the continuum form of microscopic momentum transfer.

Equation (\ref{eq:control_volume}) is the accounting tool used below. A suspended body or selected control volume can acquire momentum relative to a chosen subsystem because momentum can be transferred to the liquid, the vapor, the wall, or a wave field. A control volume can appear to gain momentum because momentum has entered through its boundary. A thermal pattern can appear to create a force because it has biased where stress is applied and where the opposite impulse is absorbed. None of this is a loophole; it is exactly how continuum momentum conservation works.

For the systems considered here, a minimal honest ledger is not simply body plus liquid. It has the schematic form
\begin{equation}
\begin{split}
\Ptot={}&\vect P_{\rm body}+\vect P_{\rm liquid}+\vect P_{\rm vapor}
+\vect P_{\rm interface}+\vect P_{\rm wall}\\
&+\vect P_{\rm acoustic}+\vect P_{\rm thermal}+\vect P_{\rm EM}
+\vect P_{\rm ionic}+\vect P_{\rm radiation}+\cdots .
\end{split}
\label{eq:ledger}
\end{equation}
The ellipsis is a checklist, not a license for speculation. Every omitted term is a possible residual channel, and every proposed channel must eventually be given a scaling law, a sign rule, a null test, and an uncertainty budget.

This is why water is a rich but demanding medium. A water-based liquid--vapor system can support surface-tension gradients, wetting transitions, contact-line hysteresis, evaporation and condensation, cavitation, capillary waves, acoustic modes, double layers, ionic screening, thermal fluctuations, and boundary compliance. Each effect is conventional. The difficulty is that several can be weak, phase-shifted, spatially distributed, or removed by averaging. A bulk-only model may miss wall energetics. A sharp-interface model may miss the finite region where stress and phase change overlap. A homogenized model may average away the very covariance that produces drift.

\section{Rectification at a liquid--vapor interface}

The minimal mechanism is covariance-driven rectification. Let $M(y)$ denote a local response property of a wall or interfacial region: mobility, wetting strength, slip length, compliance, dipole density, or another property controlling how strongly the interface responds at lateral coordinate $y$. Let $f(y)$ be a zero-mean tangential forcing. The mean forcing alone may vanish, but the correlated response need not:
\begin{equation}
\avg{Mf}-\avg{M}\avg{f}\neq0.
\end{equation}
In a two-point model the statement is exact,
\begin{equation}
\avg{Mf}-\avg{M}\avg{f}
=\frac{1}{4}(M_a-M_b)(f_a-f_b).
\label{eq:twopoint_covariance}
\end{equation}
The algebra is elementary, but the experimental consequence is easy to miss. A model that averages $M$ and $f$ separately predicts no transport; a resolved model that keeps their product predicts a directed response. If the response is uniform, the effect vanishes. If the forcing is uniform, the effect vanishes. If the relative phase is inverted, the sign can change. These nulls and reversals are the falsification handles of the mechanism.

A sinusoidal version makes the selection rule explicit. For
\begin{equation}
M(y)=\overline M+\widehat M\sin(qy+\varphi),\qquad f(y)=f_1\sin(qy),
\end{equation}
one obtains
\begin{equation}
\avg{Mf}-\avg{M}\avg{f}=\frac{1}{2}\widehat M f_1\cos\varphi,
\label{eq:sin_covariance}
\end{equation}
up to the convention used to define the phase of the drive. A phase-shifted drive gives the equivalent sine law. The invariant content is that the response is odd or even under well-defined phase operations and has exact nulls when the correlation is removed.

A concrete liquid--vapor realization is supplied by a diffuse-interface model. Let $\phi(\vect r,t)$ be an order parameter with $\phi=\pm1$ in the two bulk phases and free energy
\begin{equation}
F[\phi]=\int_{\Omega}\left[\frac{A}{4}(\phi^2-1)^2+
\frac{\kappa}{2}|\nabla\phi|^2\right]\dd V
+\int_{\partial\Omega}\left[-h(y)\phi+\frac{c}{2}\phi^2\right]\dd S .
\label{eq:free_energy}
\end{equation}
The wall imposes a Robin boundary condition through the wetting strength $h(y)$. A flat interface has profile $\phi_0=\tanh[x/(\sqrt2\xi)]$, with $\xi=\sqrt{\kappa/A}$ and surface tension
\begin{equation}
\gamma=\frac{2\sqrt2}{3}\sqrt{A\kappa}.
\end{equation}
A tangential temperature variation changes the surface tension and produces a Marangoni traction localized near the interface,
\begin{equation}
\vect f_M=\gamma_T(\vect I-\vect n\vect n)\nabla T\,\delta_\xi,
\label{eq:marangoni}
\end{equation}
where $\gamma_T=\dd\gamma/\dd T$ and $\delta_\xi$ is a diffuse-interface kernel. This is the standard tangential continuum-surface-force term, $\vect f_M=(\nabla_s\gamma)\,\delta_\xi$. For a normal fluid $\gamma_T<0$, so the traction pulls liquid from hot toward cold, toward the region of higher surface tension.

Momentum is relaxed through Stokes or Darcy--Brinkman coupling to a wall or substrate. In the Darcy--Brinkman realization, the volumetric friction $-(\mu/K_{\rm Br})\vect v$ represents momentum exchange with a porous or micro-textured substrate, such as a grooved, pillared, or coated wall wetted by the film. For a molecularly smooth solid wall, the same reservoir is reached through the boundary shear stress $\mu\,\partial_n v$ instead, and only the impedance of the wall channel changes. This reservoir is essential: without it the model would describe internal circulation, not sustained momentum transfer to a measured body.

For a weak wall heterogeneity and a zero-mean thermal drive,
\begin{equation}
h(y)=h_0+\eps_h\sin(qy+\varphi),\qquad T(y)=\Theta\sin(qy),
\end{equation}
the induced response field can be written as
\begin{equation}
M[h(y)]=M_0+\widehat M\sin(qy+\varphi)+O(\eps_h^2),
\qquad
\widehat M=\eps_h M_h,\quad
M_h=\left.\frac{\partial M}{\partial h}\right|_{h_0}.
\label{eq:mobility_amplitude}
\end{equation}
Thus $\widehat M$ is the mobility-modulation amplitude, whereas $M_h$ is the response per unit wall-field amplitude. Perturbation theory then gives the leading drift law
\begin{equation}
U=\frac{1}{2}\widehat M\gamma_T q\Theta\sin\varphi
=\frac{1}{2}\eps_h M_h\gamma_T q\Theta\sin\varphi
+O(\eps_h^2\Theta,\eps_h\Theta^2).
\label{eq:driftlaw}
\end{equation}
For a normal fluid ($\gamma_T<0$) and $\varphi=+\pi/2$ the drift is negative: the film is thickest where the cold-directed traction points backward, so the less-obstructed backward flow dominates, consistent with a lubrication estimate of the same geometry.
Thus
\begin{equation}
U(\eps_h=0)=0,\qquad U(\Theta=0)=0,\qquad U(-\varphi)=-U(\varphi),\qquad U(-\Theta)=-U(\Theta).
\end{equation}
The numerical calculations check this law beyond a collective-coordinate reduction. The full linearized phase-field problem, with the Robin wall condition, diffuse Marangoni forcing, and incompressible Darcy--Brinkman flow, recovers the same pure phase law and closes the mechanical budget against substrate drag.

\begin{figure}[t]
\centering
\includegraphics[width=0.88\linewidth]{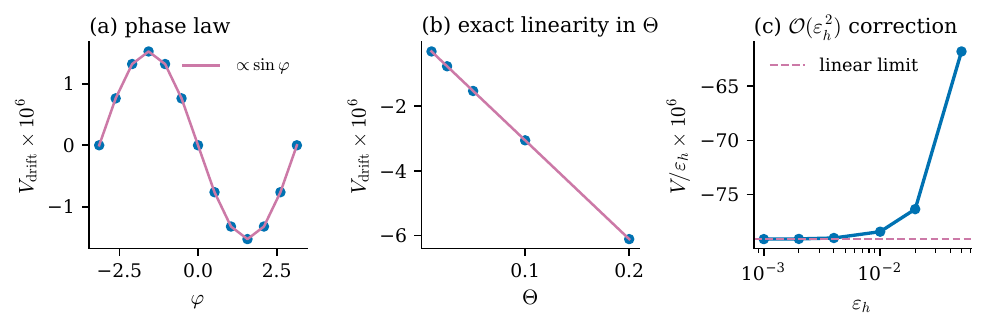}
\caption{Numerical drift law. The phase dependence is a pure sine, the response is linear in the thermal amplitude, and the heterogeneity amplitude shows the expected leading linear behavior with higher-order corrections. The point is not merely that a drift appears, but that it appears with nulls, sign reversals, and an explicit momentum sink.}
\label{fig:driftlaw}
\end{figure}

Three-dimensional textures add another useful diagnostic, not because transverse thermocapillary flow is new, but because its symmetries are sharply testable. Flow driven by transverse temperature gradients over grooved superhydrophobic surfaces has already been solved and used as microfluidic-pumping prior art \cite{Baier2010,YarivCrowdy2019,Crowdy2023}. The narrower statement here is a selection rule: for chiral wall wavevectors $\vect q_1=(a,c)$ and $\vect q_2=(b,-c)$, a spatially structured thermal drive with $\vect q_T=\vect q_1+\vect q_2$ can produce a signed transverse component,
\begin{equation}
\avg{f_z}\propto c(b^2-a^2)\cos(\varphi_1+\varphi_2).
\label{eq:chirality}
\end{equation}
The transverse component vanishes for one-dimensional patterns ($c=0$), vanishes for mirror-balanced patterns ($a=b$), and reverses under mirror inversion. This quadratic chirality factor belongs to the spatially structured-drive geometry above. A uniform AC thermal rocking protocol obeys a different selection rule: a purely AC interface response gives zero after time averaging, while a static base deformation plus AC response gives a linear chirality factor proportional to $c(a-b)\Gamma/(\Gamma^2+\Omega^2)$. The shared nulls are therefore not sufficient to identify the channel; the scaling away from the nulls and the frequency dependence distinguish the mechanisms. Such selection rules are particularly valuable in force-balance or tracer measurements, because they separate a geometrically controlled channel from uncontrolled drift.

\begin{figure}[t]
\centering
\includegraphics[width=0.88\linewidth]{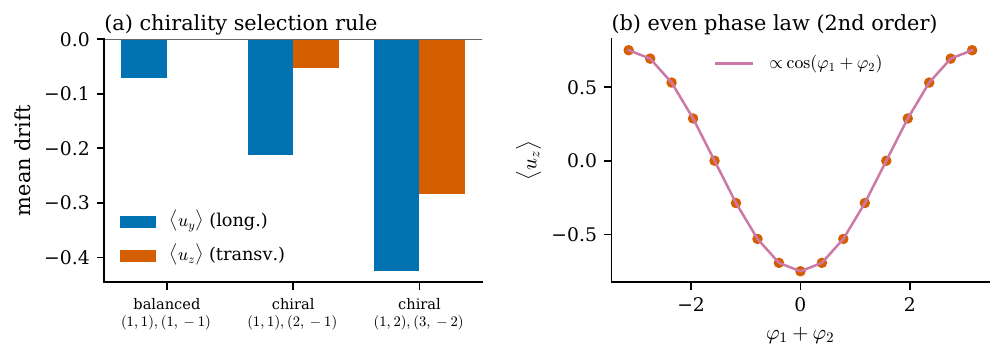}
\caption{Transverse rectification in chiral patterns. A balanced pattern nulls the transverse channel, while chiral patterns produce the predicted transverse response with an even phase law.}
\label{fig:transverse}
\end{figure}

The numerical extension reported in Appendix~\ref{app:numerics} follows the same discipline. The calculations do not attempt a complete apparatus simulation or a molecular model of water. They use a periodic spectral Brinkman solve for diffuse Marangoni forcing, a stochastic saddle-node normal form for the spinodal operating window, and a separate two-dimensional chiral-pattern calculation for transverse selection rules. Five diagnostics are reported: the sine drift law and nulls, a finite spinodal operating peak, mechanical budget closure, mirror and null tests for the three-dimensional transverse channel, and grid convergence. These calculations test whether the proposed selection rules survive a reduced flow model; they do not convert the theory into an apparatus-level claim.

\section{Susceptibility, wetting, and bounded gain}

The covariance supplies a directed forcing, but the magnitude of the response is controlled by how easily the interface moves. For a mode with wave number $q$, the useful schematic susceptibility is
\begin{equation}
\chi_q=\frac{D}{\gamma q^2+K_X},
\label{eq:susceptibility}
\end{equation}
where $D$ is the drive that actually reaches the interface, $\gamma q^2$ is the capillary penalty, and $K_X$ is the curvature of the binding potential for the interface position. This equation links covariance to gain. Soft interfacial modes can amplify rectification, but softness alone is not enough: the drive must still couple to the mode, and a reservoir must still carry the opposite momentum.

Near a first-order wetting spinodal, the bound interfacial state disappears through a saddle-node. In normal form,
\begin{equation}
V(\ell)=\frac{\ell^3}{3}-\Delta\ell,
\end{equation}
so the curvature and barrier scale as
\begin{equation}
K_X=2\sqrt{\Delta},\qquad \Delta E=\frac{4}{3}\Delta^{3/2}.
\label{eq:barrier}
\end{equation}
Mean-field susceptibility therefore grows sharply as the spinodal is approached. This is the regime in which the model predicts useful amplification: the interface is soft while still coupled to the wall. But the gain is not infinite. Thermal activation, defects, finite size, nonlinear response, and depinning round the ideal singularity and organize the response around a finite operating peak.

Critical wetting teaches the opposite lesson. There the interface unbinds continuously and the position susceptibility can diverge. One might expect unbounded amplification, but for short-range wall fields the drive reaching the interface decays as the interface unbinds. The handle disappears while the mode softens. The optimal gain therefore saturates rather than diverges. The practical distinction is between susceptibility and usable coupling: a degree of freedom can be very soft and yet poorly driven by the boundary.

The bulk liquid--vapor critical point is again different. As the bulk critical point is approached, the interface itself dissolves. Let $t=|T_{\rm abs}-T_c|/T_c$ denote the reduced distance to the critical temperature. In the self-similar protocol of the source calculations, the Marangoni rectification channel is suppressed as
\begin{equation}
K\sim t^{3\nu-1},\qquad \nu\simeq0.630,
\end{equation}
so not every criticality is a gain medium. The exponent follows from the ingredients of that protocol: $\gamma\sim t^{2\nu}$ gives $\gamma_T\sim t^{2\nu-1}$, the pattern is rescaled so that $q\xi$ remains fixed, and the normalized interfacial response overlap scales as $I\sim\xi^{-1}\sim t^\nu$. Since $K\propto\gamma_T(\xi q)I$, these factors give $K\sim t^{3\nu-1}$. Long-range van der Waals, electrostatic, or ionic couplings can extend the range over which a wall communicates with an interface, but they do not change the conservation logic. They modify the coupling $D$ in Eq. (\ref{eq:susceptibility}); they do not remove the need for a compensating momentum reservoir.

\begin{figure}[t]
\centering
\includegraphics[width=0.95\linewidth]{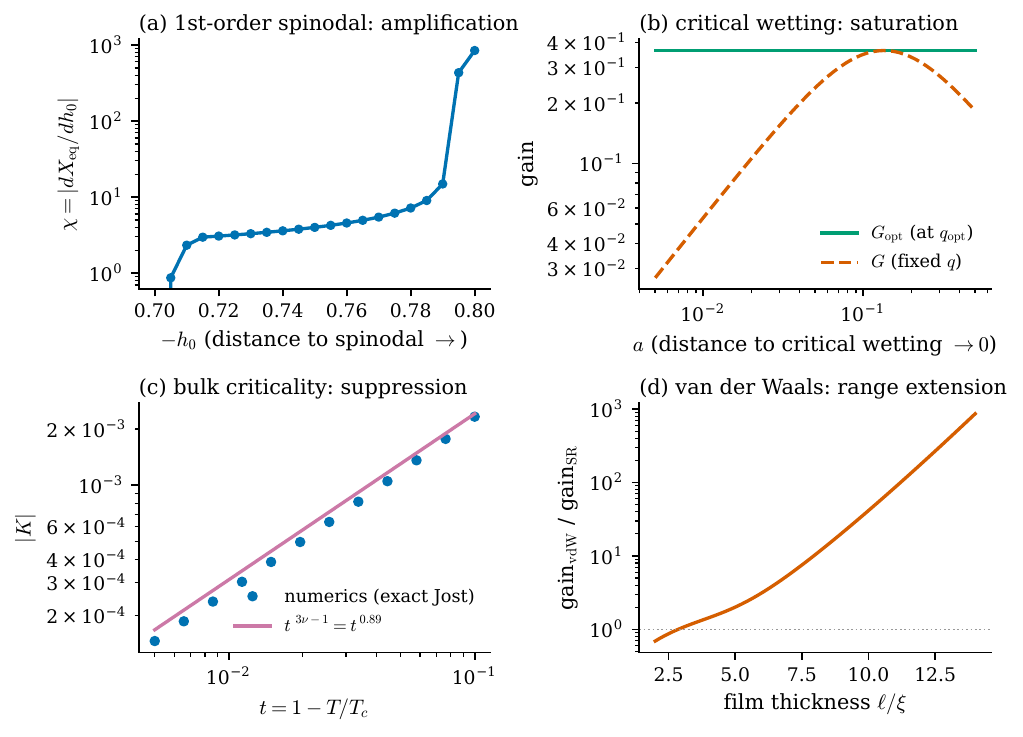}
\caption{Gain ranking from the numerical calculations. Exploitable amplification is organized by a finite peak near first-order wetting spinodal conditions; critical wetting saturates; the bulk critical point suppresses the Marangoni channel; and van der Waals tails extend the range without changing the basic accounting.}
\label{fig:ranking}
\end{figure}

Noise and compliance sharpen the same conclusion. Near a saddle-node, activated escape over the barrier in Eq. (\ref{eq:barrier}) imposes a cap. The source calculations give
\begin{equation}
\chi_{\max}\propto (k_B T_{\rm abs})^{-1/3}.
\label{eq:noisecap}
\end{equation}
Indeed, the saddle-node barrier scales as $\Delta E\propto\Delta^{3/2}$ while $\chi\propto\Delta^{-1/2}$; requiring $\Delta E\gtrsim n k_B T_{\rm abs}$ yields the one-third exponent. This thermal cap is not a loss of conservation; it is the stochastic limit on operating too close to the spinodal. Conversely, noise can enable sub-threshold transport in an asymmetric landscape, producing an interior optimum in noise strength. A compliant boundary adds another controlled susceptibility. In a reduced elastic-wall model, compliances add in series,
\begin{equation}
\chi_{XX}=\frac{1}{K_X}+\frac{1}{k_w},
\end{equation}
while stationary barrier heights remain unchanged. The wall then becomes not only a source of gain but also an explicit momentum and dissipation channel.

\begin{figure}[t]
\centering
\includegraphics[width=0.88\linewidth]{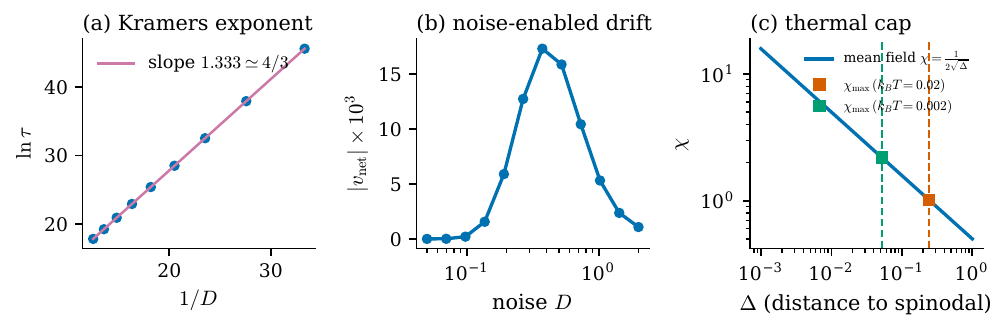}
\caption{Noise as both cap and fuel. Kramers escape caps spinodal amplification, while sub-threshold transport can have an interior optimum in noise strength.}
\label{fig:noise}
\end{figure}

\section{Additional reservoirs and experimental discrimination}

The minimal Marangoni/wetting model is intentionally clean. A real water-based system may activate additional reservoirs, and these reservoirs are not secondary details. They are precisely where an apparent momentum residual may hide.

Phase change is the most obvious extension. Evaporation and condensation introduce mass exchange, latent heat, vapor momentum, pressure transients, and recoil. If a source term $S(\vect r,t)$ is correlated with a chemical-potential or temperature pattern, it can rectify transport through the same covariance structure,
\begin{equation}
\avg{\mu S}-\avg{\mu}\avg{S}\neq0.
\end{equation}
The compensating impulse may then reside in vapor flow, pressure waves, heat flow, or the wall. In a force-balance experiment, phase change is therefore not a nuisance to be assumed away; it is a channel to be measured, suppressed, or deliberately controlled.

A scale comparison makes this discipline concrete for the vapor channel. Evaporation at mass flux $J$ exerts a normal recoil stress
\begin{equation}
p_{\rm rec}=J^2(\rho_v^{-1}-\rho_l^{-1})\simeq J^2\rho_v^{-1}.
\end{equation}
while the tangential Marangoni traction driving the ledger scales as $\tau_M\sim|\gamma_T|\,q\,\Theta$. For water at room temperature ($\rho_v\simeq2.3\times10^{-2}\,\mathrm{kg\,m^{-3}}$, $|\gamma_T|\simeq1.5\times10^{-4}\,\mathrm{N\,m^{-1}K^{-1}}$), gentle evaporation into near-saturated ambient air, $J\lesssim10^{-4}\,\mathrm{kg\,m^{-2}s^{-1}}$, gives $p_{\rm rec}\lesssim5\times10^{-7}$~Pa. This is negligible compared with $\tau_M\sim10^{-1}$--$10^{0}$~Pa for a millimetric thermal pattern with $\Theta\sim1$~K.

At reduced pressure the hierarchy reverses. The kinetic (Hertz--Knudsen) limit $J_{\max}=\alpha\,p_{\rm sat}\sqrt{M_w/2\pi R T}\simeq3\,\mathrm{kg\,m^{-2}s^{-1}}$ yields $p_{\rm rec}\sim10^{2}$--$10^{3}$~Pa, which dominates every tangential scale in the problem. Isolating the Marangoni channel in a reduced-pressure cell therefore requires operating sealed near saturation, where the net flux $J\propto(p_{\rm sat}-p_\infty)$ is suppressed, and verifying $J^2/\rho_v\ll|\gamma_T|q\Theta$ directly from the measured mass loss. Because recoil is a normal stress, it enters the tangential ledger only through its correlation with interface tilt or structure - the same covariance rule that governs every other channel.

Acoustic and capillary-wave channels provide another route. Oscillatory interfacial forcing can radiate waves, and waves carry momentum. A signal that depends on cavity geometry, acoustic damping, resonance frequency, boundary compliance, or fluid depth cannot be interpreted without including wave momentum and dissipation. Thermal channels are similarly broad: surface-tension gradients, thermophoresis, anisotropic heat leakage, buoyancy in nonideal setups, evaporation gradients, and radiative heat loss can all bias the stress field. Electrostatic and ionic channels add still another scale. Pure water self-ionizes and supports screening; patterned dipolar or ionic fields can extend wall--interface coupling beyond molecular lengths, but Debye screening limits the range. In the reduced model, pure water at room temperature gives $\lambda_D\simeq0.96\,\mu\mathrm{m}$, while millimolar salt reduces the scale to order $10\,\mathrm{nm}$; the range is therefore engineerable but not arbitrary.

These examples motivate a strict falsification discipline. A proposed channel must predict not only a magnitude, but a pattern of nulls and reversals. If the drift is proportional to $\sin\varphi$, phase reversal must reverse it. If transverse drift is chiral, mirror inversion must reverse or remove it. If the channel requires an interface, immobilizing or eliminating that interface must strongly alter the signal. If it requires evaporation, the signal must scale with mass loss, humidity, temperature, or pressure in the predicted way. If it requires ionic screening, salt concentration should move the length scale. If it requires compliance, changing boundary stiffness should change the response.

Budget closure is the final requirement. Work input, dissipation, momentum flux, and reservoir response must close within uncertainty. A residual is allowed only as a named and measurable residual, not as an absorbed modeling error. This protects both sides of the investigation. It prevents premature claims, but it also prevents real subtle channels from being dismissed merely because their momentum reservoir is distributed or unfamiliar.

The analytical and numerical checks follow the same logic. We verify the two-point covariance identity, drift-law sign reversals, saddle-node curvature and barrier scaling, critical-wetting saturation under short-range drive cancellation, chirality null conditions, and numerical recovery of the sine phase law. The reduced hydrodynamic calculations then test drift-law amplitude and nulls, finite usable spinodal gain, power-dissipation closure, chiral mirror reversal, and grid convergence. Algebraic claims are checked symbolically, selection rules logically, scaling laws numerically where possible, and empirical interpretations are tied to falsifiable protocols.

\section{Conclusions}

Four conclusions follow. First, covariance-driven rectification provides a compact way to describe directed interfacial transport from zero-mean forcing. The key object is the overlap between a structured response and a structured drive, not the mean drive alone. In the liquid--vapor realization considered here, this produces a Marangoni drift law with exact phase reversals and symmetry nulls. Those nulls are diagnostic, not cosmetic.

Second, wetting susceptibility can amplify rectified transport, but only within a bounded and physically specific regime. A first-order wetting spinodal is favorable because the interface softens while remaining coupled to the wall. Critical wetting does not provide unbounded gain for short-range wall fields because the interface unbinds and the drive reaching it vanishes at the same time. The bulk critical point suppresses the Marangoni channel as the interface dissolves. Thus the useful design principle is not ``operate near any critical point'', but rather ``operate where susceptibility grows while coupling and budget closure remain controlled''.

Third, every rectified force or drift must be assigned a compensating momentum channel. In water-based interfacial systems the channel may be liquid flow, vapor flow, wall stress, interfacial stress, acoustic or capillary waves, thermal flux, electro-ionic structure, radiation, compliance, or a combination of these. Treating these terms explicitly does not weaken the rectification claim; it makes it testable. Each channel carries its own phase law, scaling law, null experiment, and budget signature.

Fourth, the appropriate status of the model is theoretical and diagnostic. It does not assert an apparatus-level mechanism and it does not justify extrapolation beyond characterized regimes. It provides a procedure: enlarge the system, write the momentum ledger, compute the covariance and susceptibility, identify the reservoir that carries the opposite impulse, and run the symmetry and scaling controls that could falsify the interpretation. A signal that survives this process becomes an identified conventional mechanism. A signal that does not survive becomes either an artifact or an incomplete model. Both outcomes are scientifically useful.

The practical contribution is therefore a synthesis rather than a loophole: structured liquid--vapor interfaces can rectify zero-mean forcing through covariance between response and drive; wetting physics ranks where the gain can and cannot be exploited; and momentum conservation fixes the accounting discipline required to interpret the resulting transport.

\section*{Data Availability}

The numerical data and code supporting the numerical figures and consistency checks in this article are openly available in the versioned Zenodo archive ``Numerical Code for Rectified Momentum Channels,'' version 1.0.0, DOI: 10.5281/zenodo.21711829~\cite{BolivarCode2026}. The archive contains the reduced calculation code, fixed Python dependencies, reference results, and automated tests required to reproduce the reported numerical results.

\appendix
\section{Numerical methods and consistency checks}
\label{app:numerics}

The consistency checks are organized at four levels: algebraic identities, logical selection or null rules, numerical scaling, and empirical interpretation linked to a falsifiable protocol. They cover the covariance identity, sinusoidal covariance law, drift sign reversals and nulls, saddle-node scaling, short-range critical-wetting saturation, chirality nulls $c=0$ and $a=b$, and a numerical sine-fit recovery of the phase law.

The reduced hydrodynamic calculations use a periodic spectral Brinkman solver, a stochastic saddle-node normal form, and a chiral-pattern calculation. The implementations, fixed dependencies, tests, and reference outputs are archived in Ref.~\cite{BolivarCode2026}. Table~\ref{tab:numerical_checks} summarizes five representative tests. The numbers assess the internal consistency and numerical convergence of the reduced model; they are not experimental calibration.

\begin{table}[t]
\caption{Numerical consistency checks for the reduced hydrodynamic model: selection rules, finite spinodal optimum, budget closure, three-dimensional chirality signs, and convergence.}
\label{tab:numerical_checks}
\centering
\small
\begin{tabular}{p{0.25\linewidth}p{0.65\linewidth}}
\toprule
Test & Numerical result \\
\midrule
Drift law and nulls & Resolved Brinkman calculation recovers $U\propto\eps_h\Theta\sin\varphi$ with $R^2=1.0$; maximum absolute prediction error $4.24\times10^{-21}$; $\eps_h=0$ and $\Theta=0$ nulls pass. \\
Spinodal optimum & Stochastic saddle-node model has a finite interior peak at $\Delta=4.73\times10^{-2}$ with usable gain proxy $6.53\times10^{-2}$, rather than an exploitable divergence. \\
Mechanical budget & Input power and dissipation close: $P_{\rm in}=1.70762359264151\times10^{-6}$, $D=1.7076235926415103\times10^{-6}$, relative residual $2.48\times10^{-16}$. \\
3D chiral channel & Transverse calculation gives $\tan\theta=0.25$; mirror $z$-sum $5.55\times10^{-16}$; balanced-pattern and one-dimensional nulls pass. \\
Grid convergence & Finest grid $N=256$; finest budget residual $2.48\times10^{-16}$; residual decay factor $1.44\times10^{7}$. \\
\bottomrule
\end{tabular}
\end{table}

\end{document}